# Dynamic friction force in a carbon peapod oscillator

**Haibin Su[1,2], William A. Goddard III[1], Yang Zhao[3]**

[1])Materials and Process Simulation Center, California Institute of Technology, Pasadena, CA 91125, U.S.A.
[2])Division of Materials Science, Nanyang Technological University, Singapore
[3])Department of Chemistry, University of Hong Kong, Pokfulam Road, Hong Kong, China

Abstract

We investigate a new generation of fullerene nano-oscillators: a single-walled carbon nanotube with one buckyball inside with an operating frequency in the tens-of-gigahertz range. A quantitative characterization of energy dissipation channels in the peapod pair has been performed via molecular dynamics simulation. Edge effects are found to the dominant cause of dynamic friction in the carbon-peapod oscillators. A comparative study on energy dissipation also reveals significant impact of temperature and impulse velocity on the frictional force.

Nanoscale fabrication technologies have made pervasive impact in the past 20 years[1]. One of the manifestations has been in the area of nano-electro-mechanical-systems (NEMS)[2], which broadly refers to the application of nano-fabrication technologies to construct sensors, actuators, and nano-scale integrated systems for a variety of applications. Very recently, Zettl's group reported that frictional forces are very small, c.a. in the magnitude of $10^{-14}$ N per Å$^2$, during the controlled and reversible telescopic extension of multi-wall carbon nanotubes[3]. Furthermore, it has been proposed that the transit time for complete nanotube core retraction (on the order of 1 to 10 ns) implies the possibility of exceptionally fast electromechanical switches[4]. In fact, oscillating crystals have a long history, dating back to 1880 when the piezoelectric effect was discovered by the Curie brothers[5]. Quartz crystal oscillators are widely used to provide regular pulses to synchronize various parts of an electronic system. But a typical crystal is millimeters in size, which could not be directly integrated into a computer chip. Motivated by the observation of Zettl's group's, Zheng and Jiang[4] proposed a new type of nano-oscillators operating completely differently from conventional quartz oscillators. Since then, designing this type of nano-oscillator has been carried out actively. Legoas and collaborators[6] first simulated an 38-GHz nano-oscillator consisting of a (9,0) carbon nanotube (CNT) inside of an (18,0) CNT. Zhao *et al*[7] found that off-axial rocking motion of the inner nanotube and wavy deformation of the outer nanotube are responsible for energy dissipation in a double-walled nanotube oscillator.

So far, no successful experimental realization of the bi-tube oscillators has been reported. This is probably due to the considerable amount of energy dissipation, and the difficulty of preparing bi-tube type oscillator unit from multi-wall carbon nanotubes with high quality. Fortunately, we have effective ways to place buckyballs inside nanotubes. For instance, single-walled carbon nanotube (SWNT) can be synthesized by pulsed-laser vaporization route, whereby the sublimation of solid $C_{60}$ in the presence of open SWNT causes the fullerenes to enter SWNT and self-assemble into 1D chains[8]. Therefore, it is feasible for single $C_{60}$ to enter nanotube by van der Walls interactions. The peapod formation process has been widely studied[9-12]. Many studies are also focused on the effect of the nanotube diameter on binding properties[9, 10]. Filling SWNT with $C_{60}$ is exothermic or endothermic depending on the size of the nanotube. $C_{60}$@(10,10) is found to be stable (exothermic) while other peapods with smaller radius such as the (9,9) and (8,8) tubes are endoethermic[10]. Among many proposed interesting applications of peapod structures, one recent nanomechanical resonance study has been performed on C60-filled carbon nanotube bundles towards developing next generation resonating systems[13]. The topic of this article is another feasible application of carbon-peapod system. Comparing with bi-tube structure models, here we propose that it is much more practical to replace the inner tube with a buckyball since the dynamic friction force between two objects is indeed proportional to the area of the overlapping sections from the perspective of modern tribology. Similar carbon-peapod osillators were investigated previously[14, 15], focusing on elastic properties[14] and length-dependent oscillating frequencies[15]. It was also found that the frictional behavior of the carbon peapod depends on the diameter and chirality of the nanotube[15]. However, little attention has been paid so far to detailed energy dissipation mechanisms in the carbon-peapod oscillators. It is the aim of this article to investigate energy dissipation channels and effects of temperature and impulse velocity in the peapod oscillators via molecular dynamics simulation.

The structure model consists of one (10,10) SWNT of length 50.05Å with one $C_{60}$ molecule inside. The edges of SWNT are passivated with hydrogen atoms. The diameter of $C_{60}$ is 6.83Å, which fits nicely into (10,10) tube with a diameter 13.56Å (see Fig. 1). The force field used here for $sp^2$ carbon centers was developed by fitting experimental lattice parameters, elastic constants and phonon frequencies for graphite[16]. This force field uses Lennard-Jones 12-6 van der Waals interactions ($R_v$ = 3.8050, $D_v$ = 0.0692), Morse bond stretches ($R_b$ = 1.4114, $k_b$ = 720, $D_b$ = 133.0), cosine angle bends ($\theta_a$ = 120, $k_{\theta\theta}$ = 196.13, $k_{r\theta}$ =-72.41, $k_{rr}$ = 68), and a twofold torsion ($V_t$ = 21.28), where all distances are in Å, angles in degrees, energies and force constants in kcal/mol. This force field correctly predicts that the energetically favorable packing for $C_{60}$ is face-centered cubic and for $C_{70}$ is hexagonal close-packing (h.c.p.). A more interesting result relevant to this work is the good agreement of the sublimation energy between calculation (40.9 kcal/mol at 739 K) and measurement (40.1 $\pm$ 1.3 kcal/mol at 739 K)[16]. This force field is employed here to study the binding energy $E_b$

$$E_b = E_{C_{60}@(10,10)} - E_{(10,10)} - E_{C_{60}} \qquad \text{(Eq. 1)}.$$

It is expected that buckyball molecules are more attracted to a SWNT than to each other due to a larger contact area with SWNTs, and therefore, more carbon-carbon van der Waals interactions. Once a fullerene enters a nanotube, the van der Waals attraction keeps it inside. From our force-field parameters, the binding energy is – 81.4 kcal/mol for putting one $C_{60}$ inside a (10,10) SWNT, which is consistent with previous results reported by the Girifalco group[17] and Ulbricht et al[11]. However, the reported binding energies by Okada et al[10] (c.a. –11.8 kcal/mol) and the Louie group[12] (c.a. –23.1 kcal/mol) are different from the above results, which is not surprising as it is well-known that long-range attractive (London dispersion) interactions are not adequately described by the DFT methods[18] based on the local density approximation and the generalized gradient approximation..

As a preparation, our system is first equilibrated at 120K, 180K, 240K, 300K, and 360K with the NVT dynamics for 30 ps. After an impulse velocity is added to buckyball, dynamics simulation is carried out with the thermostat only attached to the tube. It is clear that the total energy of the $C_{60}@(10,10)$ system is not conserved. It is desirable to study the energy dissipation channels which allow energy flow from buckyball to tube, and then to the thermostat.

The tube serves as a potential well to confine the motion of the ball. It is easy to estimate the maximally-allowed impulse velocity, which is 960 m/s, using the relation $v_{max} = \sqrt{2E_b/m_{c_{60}}}$. In our simulation, the $C_{60}$ molecule is given an initial translational velocity of 480 m/s. Various impulse velocities between 100 m/s and 700 m/s also have been applied for T = 300K in order to study the relation between the dynamic friction force and the initial velocity. Kinetic energy data of the buckyball from the molecular dynamics simulation are smoothened by averaging over every five time periods (see Fig. 2a). As the buckyball moves to the opening end of the tube, it is slowed down until the *translational* velocity vanishes thanks to the van der Waals attraction between tube and ball. At this moment, the potential energy reaches its local maximum, after which the ball returns back to the tube (see Fig. 2b and 2c). The period is 20 ps, corresponding to a frequency of 50GHz, which is quite encouraging for its potential applications to the nano-fabrication field.

In this system, there are two important channels for energy dissipation: off-axial wavy motion and the edge effect. The former means that the ball moves off-axially inside the tube; the latter refers to the case when the ball moves to the edges of the tube. Macroscopic models of friction between solids dictate that friction is proportional to normal force, independent of contact area. This is so-called Amontons's law. For dynamic friction force, Coulomb's law states that it is independent of velocity. It is interesting to note that recent experiments[19-21] by atomic force microscope (AFM)[22] suggest that microscopic friction does not always behave according to traditional Amontons's law and Coulomb's law, thereby suggesting the need for new laws that account for atomic scale phenomena[23, 24]. So far, both area and velocity dependences have been demonstrated. It should be emphasized that edge effects should be paid more attention at nanoscale. This will be illustrated further in this Letter.

The dynamic friction force can be readily computed from the energy data

$$f = -\frac{\Delta E}{\Delta l},$$ (Eq. 2)

where $\Delta l$ is the distance the buckyball travel while its kinetic energy is decreased by an amount $\Delta E$. Comparing with previous studies on double-walled oscillators, here only the tube is attached to the thermostat, a setting arguably more convenient to study energy dissipation in nanotube oscillators. Under this circumstance, the simulation yields the upper bound of dynamic friction force. It is interesting to compare the dynamic friction force at 300K evaluated in this study with those from the literature. In Zettl's paper, the dynamic friction force per area is estimated to be less than 4.3 x $10^{-15}$ N/Å$^2$. In our study, this dynamic friction force is 0.17 pN, which is far less than that in bi-tube-like system (usually in the magnitude of nN). Note that the contact area between ball and tube is 97.9

Å$^2$. This means the dynamic friction force per unit area of our system is 1.8 x 10$^{-15}$ N/Å$^2$ which contains contributions from both the off-axial motion and the edge effects.

We now address the temperature effect on dynamic friction. The results, plotted in Fig. 3a and 3b, show that the dynamic friction force increases as the raising temperature. The source of this dynamic friction is the energy flow from the ball to the tube. This is caused by both the off-axial motion between the ball and the tube, and the edge effects. We define the off-axial angle, $\theta$, of buckyball with respect to the tube as

$$\theta = \cos^{-1}(\frac{abs(\vec{r}_1 \bullet \vec{r}_2)}{|\vec{r}_1| \times |\vec{r}_2|}),$$ (Eq. 3).

where $\vec{r}_1$ is the vector from origin to the center of ball, $\vec{r}_2$ is the axial vector from origin to the right edge of tube, *abs* means taking absolute value so that the value of $\theta$ is in the first quadrant (see Fig. 3c). No constraint is applied on the shape of the SWNT during the simulation. Indeed, there exists an energy transfer channel via the coupling between the off-axial wavy motion of the buckyball and the radial breathing mode of the SWNT. At the atomic scale, the shape of SWNT cannot be perfectly rigid due to thermal vibrations. Especially, when the buckyball moves near the edges of the SWNT, the axial vector of SWNT tilts away by about 0.4$^o$ from x-axis (assuming the SWNT initially is aligned along the x-axis). Thus, this variation should be taken into account. In practice, we use the mass center of the right edge of tube as the reference point to compute the axial vector, and then calculate the off-axial angle based on Eq. 3 for each ps during the simulation. The average off-axial angles tabulated in Table 1 are analyzed following this procedure. In Table (1.a), the average off-axial angles for different temperatures with a fixed impulse velocity (v = 480m/s) are collected. There is a clear trend that $\theta$ increases monotonically with the temperature, indicating that the ball takes on larger off-axial motion at higher temperatures which leads to higher energy dissipation. In addition, the energy transfer between the ball and the tube at the tube edges is also enhanced. Therefore, the dynamic friction force increases (see Fig. 3a). In this case, the contributions to dynamic friction from two dissipation channels are mixed together. However, as we stated before, when the buckyball moves near the edges of the SWNT (see Fig. 3d), there is significant energy transfer between them. Therefore, we must study quantitatively energy dissipation due to the edge effect. To do this, we have examined the relation between the dynamic friction force and the impulse velocity in this system by varying impulse velocities (between 100 m/s and 700 m/s) at T = 300K. The results are plotted in Fig. 3b. Indeed, the velocity dependence of dynamic friction has been reported by Gnecco et al[21] for a silicon tip sliding on a NaCl(100) surface. In our simulation, the velocity is in the order of $10^2$ m/s, which is ten orders of magnitude higher than those in Gnecco et al's paper[21]. Thus, we are in two completely different velocity regimes. In addition, the friction force calculated here arises from smooth sliding between a bukcyball and the inner walls of a SWNT without applied normal forces, while in Gnecco et al's paper, strong forces are applied between the tip and the NaCl (100) surface. Consequently, a direct comparison can not be made. However, good agreement has been found between this work and another earlier computational study[7] on the value of frictional forces per atom. In our model it is interesting to note that the average off-axial angles are little influenced by impulses for a given temperature, for example, 300K (see Table (1b)). Since the frictional force increases with the impulse velocity as shown in Fig. 3b, it follows that this velocity dependence is mainly due to energy dissipation at the edges of the tube (not due to off-axial motion), and the larger impulse velocity, the more energy dissipation at the edges of the tube. More importantly, we can extract the dynamic friction force due to off-axial wavy motion at 300K by simply extrapolating the data to zero impulse velocity. It yields a friction force of about 12 fN due to the off-axial wavy motion. For comparison, the contribution from the edge effects (158 fN) with an impulse velocity of 480m/s is about one order of magnitude larger than wavy motion at 300K with the same impulse velocity. This is supported by a recent independent study by Tangney et al[25].

To conclude, we have carried out a quantitative characterization of dominant energy dissipation mechanisms for a carbon peapod nano-oscillator that can be realized in the lab. Our molecular dynamics simulation results reveal significant effects of the temperature and the impulse velocity on friction. In particular, it has been shown that the edge effects are the main cause of the dynamic friction force. This novel nano-oscillator design proposed here holds great promise for applications in NEMS owing to its extremely low operating friction and easy adaption for impulse generation[26].

## Acknowledgement

H.B.S. is grateful for kind assistances in coding from Y. Lansac, J. Dodson, and P. Meulbroek. The work at NTU is funded by NTU-CoE-SUG under Grant No. M58070001. The facilities of the Materials and Process Simulation Center (MSC) used in these studies are funded by DURIP (ARO and ONR), NSF (CTS and MRI), and a SUR Grant from IBM. In addition, the MSC is funded by grants from ARO-MURI, NIH, ChevronTexaco, General Motors, Seiko-Epson, the Beckman Institute, Asahi Kasei, and Toray Corp.

[27] For instance, see J. C. Hummelen et al, Science **269**, 1554 (1995); and J. C. Hummelen et al, Journal of the American Chemical Society **117**, 7003 (1995).

# Table

**Table 1.** Temperature and velocity effects on off-axial angles. In (1.a) the impulse is fixed as 480 m/s for all the temperatures. In (1.b) the temperature is kept to be 300 K while varying impulses. The off-axial angles are averaged for 80 periods during the oscillation. See text for detailed discussions.

1.a.

| T (K) | 120 | 180 | 240 | 300 | 360 |
|---|---|---|---|---|---|
| $\theta$ (degree) | 1.54 | 1.61 | 1.90 | 2.34 | 2.50 |

1.b.

| V (m/s) | 100 | 180 | 280 | 380 | 480 | 680 |
|---|---|---|---|---|---|---|
| $\theta$ (degree) | 2.41 | 2.42 | 2.42 | 2.42 | 2.34 | 2.26 |

# FIGURE CAPTIONS

Fig. 1. (color online) Structure model of the novel oscillator. One $C_{60}$ molecule is inside (10,10) SWNT of length 50.05Å. The diameter of $C_{60}$ is 6.83Å, which fits nicely into (10,10) tube with diameter 13.56Å.

Fig. 2. (color online) Evolution of energy data from molecular dynamics simulations for impulse velocity v = 480m/s at 300K. (a) Evolution of buckyball kinetic energy in whole time domain of the simulation. The solid circles represent energy data by averaging every five periods. (b) Evolution of buckyball kinetic energy between 120 and 170 ps. (c) Evolution of potential energy of buckyball between 120 and 170 ps.

Fig. 3. (color online) Effects of temperature and velocity on dynamic friction force, and two channels for energy dissipation. (a) The impulse velocity is fixed as 480m/s. (b) The temperature is fixed as 300K. (c) Off-axial motion of $C_{60}$ relative to (10,10) SWNT, where $\vec{r}_1$ is the vector from origin to the center of ball, $\vec{r}_2$ is the axial vector from origin to the right edge of tube, and the off-axial angle, defined in Eq. (2) in the text, is exaggerated for better visualization; (d) The motion of $C_{60}$ near edges of SWNT. Please see the text for a detailed discussion.

**Haibin Su, William A. Goddard III, and Yang Zhao:    Fig. 1**

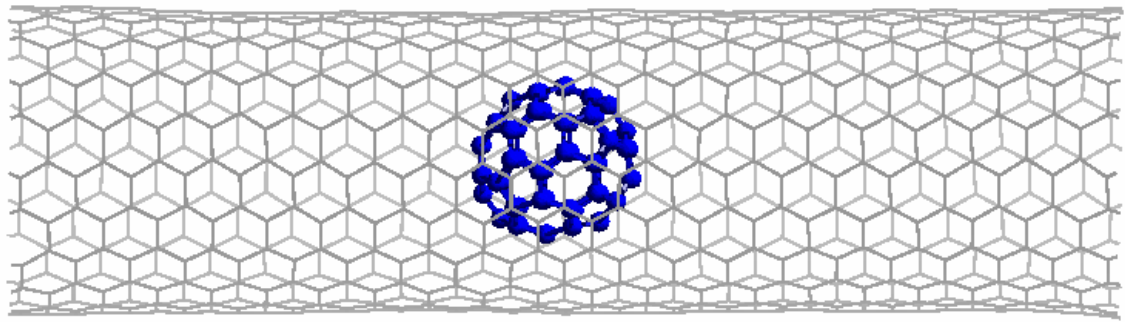

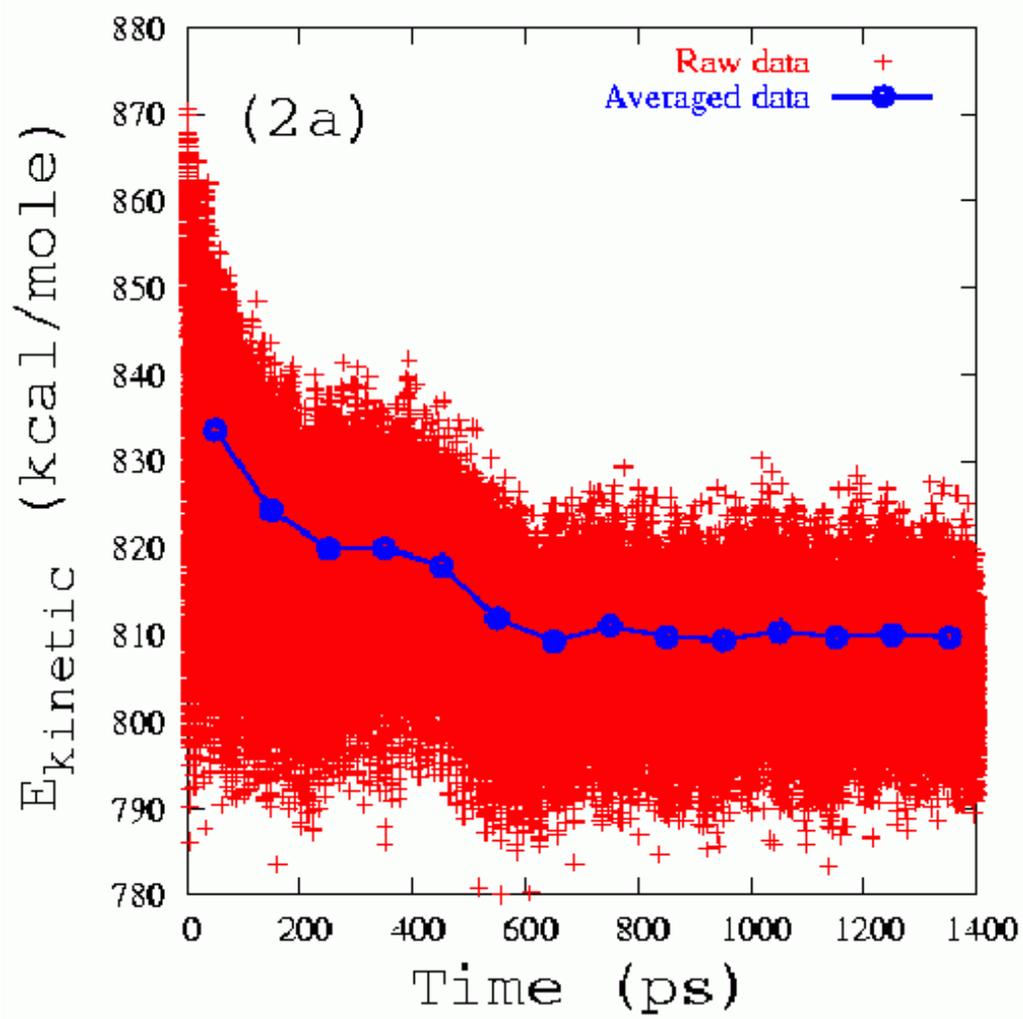

**Haibin Su, William A. Goddard III, and Yang Zhao:   Fig. 2bc**

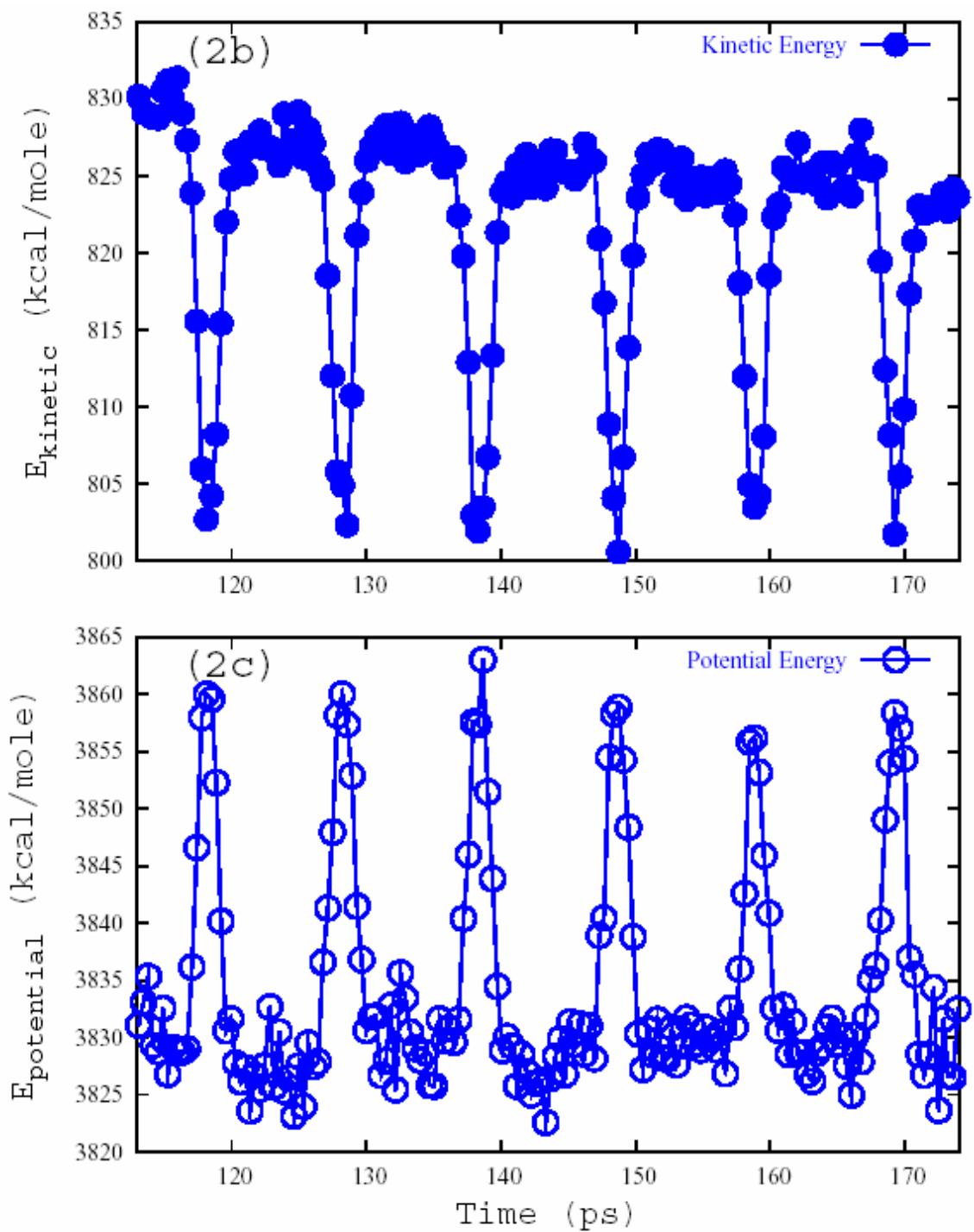

**Haibin Su, William A. Goddard III, and Yang Zhao:     Fig. 3**

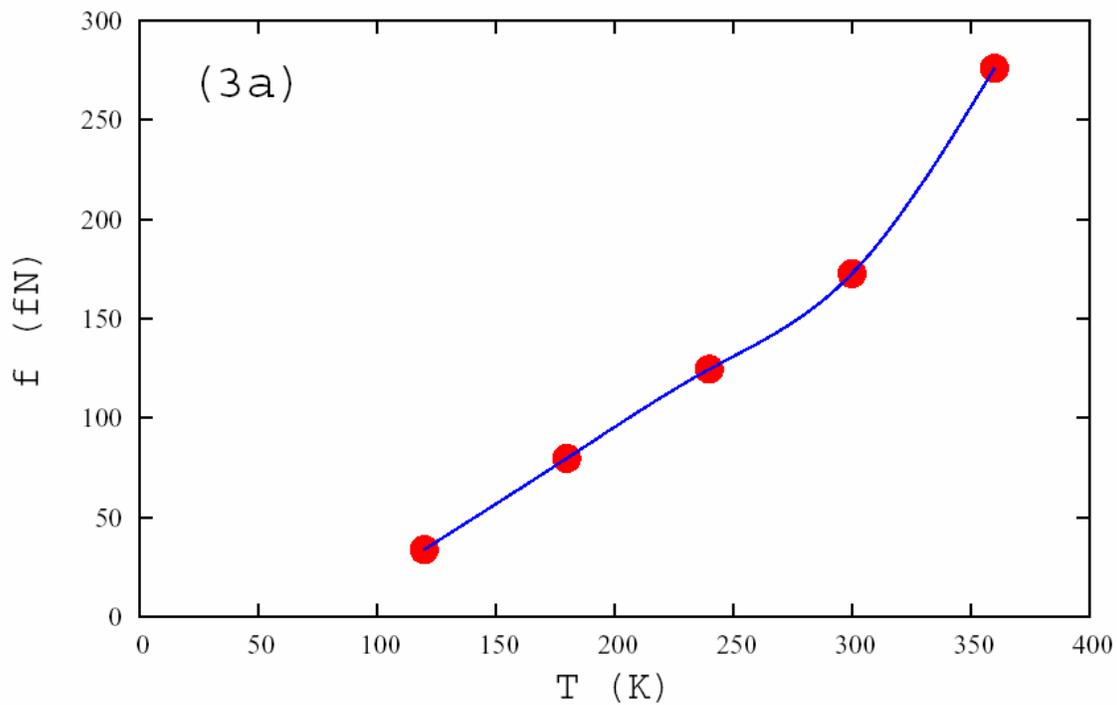

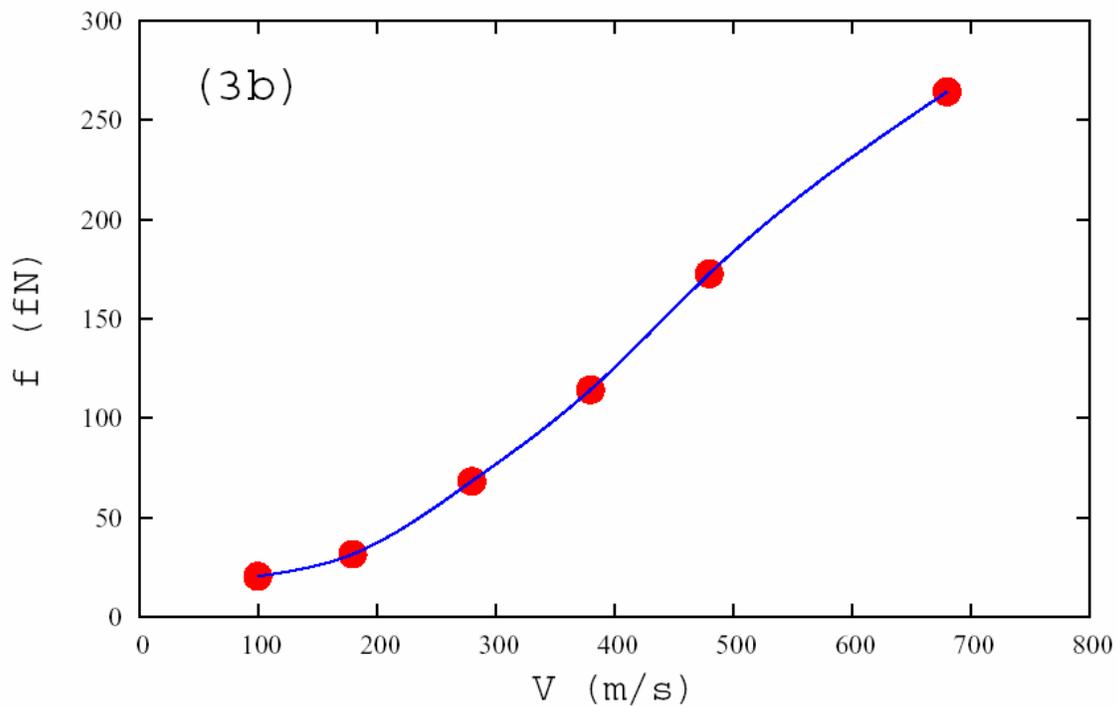

**Haibin Su, William A. Goddard III, and Yang Zhao:    Fig. 3 (continued)**

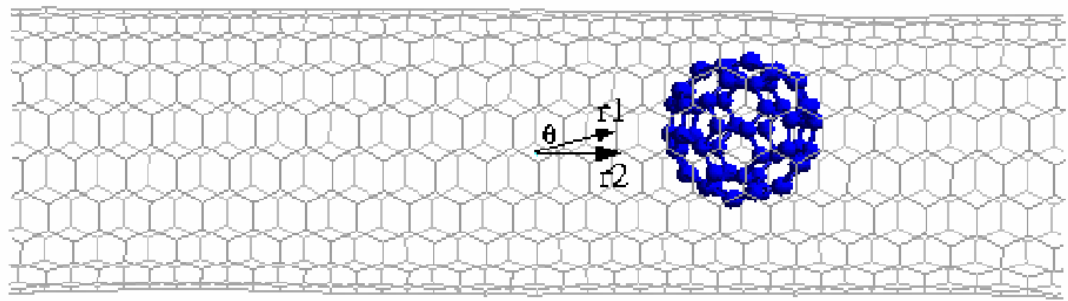

Fig. 3c

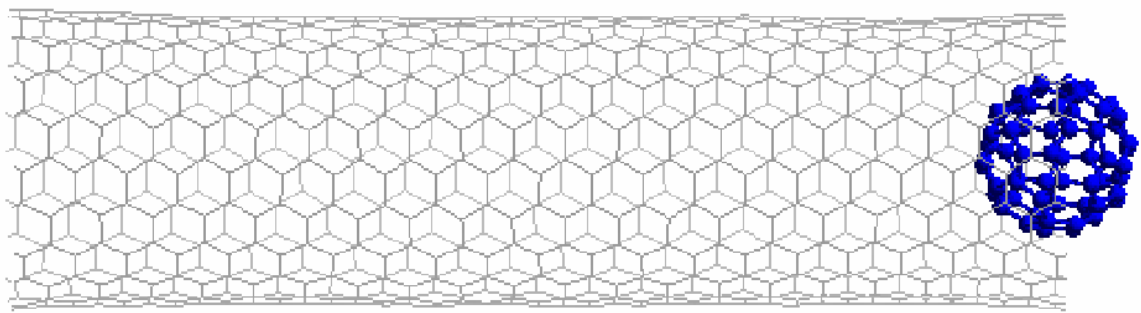

Fig. 3d